\documentclass[useAMS,usenatbib,usegraphicx]{mn2e}
\title[The helium spread in the globular cluster 47 Tuc]{The helium spread in the globular cluster 47 Tuc}
\author[]{M. Di Criscienzo$^{1}$\thanks{E-mail:dicrisci@oa-roma.inaf.it}, P. Ventura$^{1}$, F. D'Antona$^{1}$, A. Milone$^{2}$ $\&$ G. Piotto$^{2}$\\
$^{1}$INAF-Osservatorio Astronomico di Roma,
              Via Frascati 33, I-00040, Monte Porzio Catone, Rome, Italy\\
$^{2}$ Dipartimento di Astronomia,Universit\'a di Padova,
              Vicolo dell' Osservatorio 3, Padova, I-35122, Italy\\}
\begin{document}
\date{Accepted 2010 June 8.  Received 2010 May 19; in original form 2010 February 9}

\maketitle

\label{firstpage}
\begin{abstract}
Spectroscopy has  shown the presence of the CN band dicothomy and the Na-O anticorrelations for 50--70\% of the investigated samples in the cluster 47 Tuc, otherwise considered a "normal" prototype of high metallicity clusters from the photometric analysis. These  anomalies are also found the Main Sequence stars, suggesting  that a consistent second generation is present in 47 Tuc. 
Very recently, the re-analysis of a large number of archival HST data of the cluster core has been able to put into evidence the presence of structures in the Sub Giant Branch:  it has a brighter component with a spread in  magnitude by $\sim$0.06 mag  and a second one, made of about 10\% of stars, a little fainter (by $\sim$0.05 mag).  
These data also show that the  Main Sequence of the cluster has an intrinsic spread in color which, if interpreted as due to a small spread in helium abundance, suggests $\Delta$Y$\sim$0.027.\\
In this work we examine in detail whether the Horizontal Branch morphology and the Sub Giant structure provide further independent indications that a real  --although  very small-helium spread is present in the cluster.
We re--analyze the HST archival data for the Horizontal Branch of 47 Tuc, obtaining a sample of $\sim$500
stars with very small photometric errors, and build population synthesis based on new models 
to show that its particular morphology can  
be better explained by taking into account a spread in helium abundance of 2$\%$ in mass. 
The same  variation in helium  is able to explain  the spread in luminosity of the Sub Giant Branch,  
while a small part of the second generation is  characterized by a small C+N+O increase and provides 
an explanation for the fainter Sub Giant Branch.
We conclude that three photometric features concur to form the paradigm 
that a small but real helium spread is present in a cluster that  
has no spectacular evidence for multiple populations like those shown by other massive clusters. This work thus shows that multiple populations in Globular Clusters are more and more confirmed  to be ubiquitous.
\end{abstract}

\section{Introduction}
NGC 104, better known as 47 Tuc, is the second largest and brightest globular cluster (GC) in the sky and for this reason it is  one of the most studied old associations of the Milky Way. Although not so extreme  as the other two GCs of similar metallicity, NGC6441 and NGC6388, this cluster shows some chemical anomalies, that deserve investigation.
On the spectroscopic side, studies
in the last 30 years show CN variations along the Red Giant Branch (RGB)  of 47 Tuc down to the Turn Off (TO) and along the Main
Sequence (MS). In particular \citet{briley1997} verified the
existence of a bimodal distribution of CN band strengths. An examination of the location of 
Horizontal Branch (HB) stars  on the color-magnitude diagram reveals that the CN-strong HB 
stars are on average, about 0.05 mag more luminous in V, and tend to be bluer, than their 
CN-weak counterparts. 
\citet{briley1997} found also a radial gradient of the relative fraction of stars with
strong and weak CN bands. In the inner part of the cluster
there is a higher fraction of stars with strong CN absorption. The existence of this gradient had
been first noted by \citet{norrisfreeman1979} and further
documented by  \citet{pataglou1990}.
\citet{cannon1998} showed that the
bimodal distribution of CN band strengths is present also  on the
upper MS. Recently,  with the availability of 8 to
10 m class telescopes and the capability of multiobject
spectroscopy, \citet{harbeck2003}  have shown that the bimodality in CN strength still exists 2.5 mag below the TO.\\
These results indicate that the abundance spread in 47 Tuc is not due to an
evolutionary effect, rather to the presence of an original stellar population
(first generation, FG) and of a second generation (SG). Stars in the SG
formed from material processed through the hot CNO cycle in the progenitors, belonging to the FG, but not enriched in the heavy elements expected in supernova ejecta,  according to todays' new
paradigm for the formation of galactic GCs \citep{gratton2001,carretta2009a,carretta2009b,dercole2008}.\\It is not definitely settled what kind of stars  produced this material. The two most popular candidates are  the intermediate mass stars during the Asymptotic Giant Branch (AGB) phase \citep{ventura2001} or fast rotating massive stars \citep{decressin2007}, and  is not clear yet if the material came entirely from the wind ejected from these objects  or it is a mixture of processed gas and pristine matter of the initial  star forming cloud.\\
The analysis of the Na-O anticorrelation among the stars of 47 Tuc performed by  \citet{carretta2009a,carretta2009b} gives results compatible with this interpretation: about 70 \% of stars belong to what they call ``intermediate'' population, i.e. they  are oxygen depleted and sodium  enhanced.\\
Although other clusters also show evident photometric indications that multiple population are present, 47 Tuc did not show any clear photometric sign, until \citet{anderson2008}  which  exploiting the large number of archival HST data have  found, in the cluster core, a splitted sub giant branch (SGB) with at least two distinct components: a brighter one  with a small and   real spread in magnitude ($\sim$0.06 mag)  and a second one containing about 10\% of star a little ($\sim$0.05mag) fainter.\\ 
Recently Bergbusch $\&$ Stetson (2009) have published a color magnitude diagram, based on ground-based data, covering all evolutionary sequences of 47 Tuc. They also looked at the possible presence of multiple sequences, in particular for star brighter than the turn off. They could not reach
any conclusive evidence of a multiple SGB, which is not surprising, because of the lower quality of the ground-based data, and, most importantly,
because of the fact that, because of crowding, they could concentrate only on the outer part of cluster. Small number statistics, and the possible presence of a radial gradient in the multiple population distribution (as in $\omega$~Cen, Bellini et al. 2009)
may be at the basis of their failure to find the features so clearly identified in the cluster core by \citet{anderson2009}.\\
A splitting of the SGB was also observed in NGC 1851 \citep{milone2008}. \citet{cassisi2008}  explained this result with a difference in the overall C+N+O content of the two group of stars, and negligible spread in helium, whereas \citet{ventura2009}, in the framework of a self-enrichment process by which a second generation of stars formed from the gas ejected by massive AGBs, assumed that stars in the faint SGB were not only enriched in the CNO by a factor $\sim 3$, but also sligthly enriched in helium, in agreement with the yields of the intermediate masses.
If we make the same hypothesis for 47 Tuc,  a contradiction emerges because only the  10\% of stars belong to faint SGB, not enough  to explain the percentage of SG inferred by the spectroscopic data.\\

To explain the formation and constitution of 47 Tuc stars, in this work we build a homogeneous framework to reconcile the spectroscopic  and photometric resuts by \citet{anderson2009}. For this aim, we will use both HB and SGB data (Section 3), interpreted on the basis of the models described in Section 2.
A discussion of the results and conclusions close the paper.  

\section{Evolutionary models and population synthesis}
We computed  stellar models  with the code ATON2.0, described in \citet{ventura1998} and updated in \citet{ventura2009}. 
We adopted a metallicity Z = 0.006 and an $\alpha$-enhanced mixture with [$\alpha$/Fe]=0.4. As standard helium content we have adopted Y=0.25, but different initial Y (=0.28, 0.32, 0.40) were investigated. Models cover   the mass range $\sim$  0.40M$_{\odot}$-1.2M$_{\odot}$ from the pre main sequence  until the He ignition at the RGB tip. 
In the case of  HB stellar models, the He-core mass and envelope He-abundance
values were taken from the RGB progenitors for an  age of $\sim$ 12Gyr.\\
\begin{figure}
\includegraphics[width=9cm]{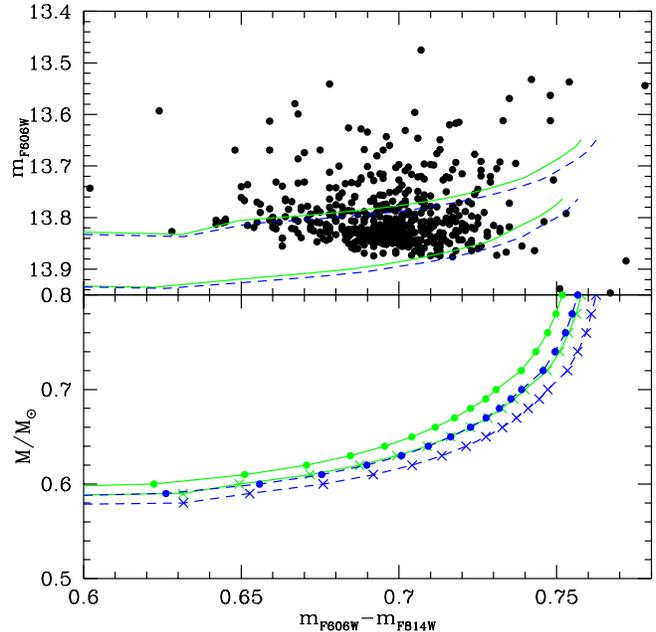} 
\vspace{-50pt}
\caption{Upper panel:CMD diagram of 47 Tuc HB stars. ZAHB are also reported for normal(solid line) and CNO$\uparrow$(dashed line) populations with respectively Y=0.25 (low luminosity) and Y=0.28 (high luminosity) shifted by m-M$_{\rm F606W}$=13.09 mag  and E(606-814)=0.038 mag (see text). Lower panel: stellar mass vs color  for  ZAHB models. Filled  circles and crosses  refer to models with Y=0.24 and Y=0.28 respectively.}
\label{fig1}
\end{figure}
In order to reproduce the faint SGB we followed the line of interpretation adopted for the double SGB in NGC1851 \citep{ventura2009}, and considered C+N+O enriched models. Concerning the choice of the element mixture, we make the hypothesis that this population is born from matter mixed with hot-CNO processed ejecta of massive AGBs and look at the abundances computed by \citet{venturadantona2009} (Table 2). Extending their results for the metallicity of 47 Tuc we see that  the 5M$_{\odot}$ AGB evolution provides a C+N+O increase by a factor $\sim$1.4, with C depleted by --0.55 dex, N enhanced by 1.44 dex   and O depleted by 0.31 dex (i.e [O/Fe]=0.09 dex for our $\alpha$-enhanced mixture). We computed evolutionary tracks and isochrones for this composition (mixture CNO$\uparrow$$\uparrow$) and also for a compositions obtained by diluting these abundances with 50\% of matter having the starting standard composition (we call this mixture CNO$\uparrow$). The idea below these choices is that  the AGB processed
material may has been  mixed with some  amount of pristine material.
Such a  dilution model has been, for example, used to
explain the features of the Na-O anticorrelation of NGC 6397 \citep{venturadantona2009,dicriscienzo2009}.\\
The helium abundance of the new mixture CNO$\uparrow$$\uparrow$ would be Y=0.31 \citep{venturadantona2009} while for the mixture CNO$\uparrow$ it will be Y=0.28. The value Y=0.31 is not compatible with the intrinsic spread  of the MS  found by \citet{anderson2009} while Y=0.28 is more acceptable.
Concerning the total ``metallicity'' Z in mass fraction we have Z=0.008 (=0.007) for CNO$\uparrow$$\uparrow$ (CNO$\uparrow$). We compute on purpose radiative opacities for this mixture using the OPAL web tool  only for T$\ge$10000K,  as low temperature opacities do not affect the structure of the models we are considering \citep{ventura2009}.\\
All these models were used to derive synthetic populations in order to reproduce the observed features of 47Tuc's CMD.
 In particular synthetic populations for the HB are built, following the prescription by 
\citet{dantonacaloi2008}. We adopt an appropriate relation between the mass of the evolving giant M$_{\rm RG}$ and the age, as function of helium content at this metallicity.
The mass on the HB is then M$_{\rm HB}$ = M$_{\rm RG}$(Y,Z)--$\Delta$M, where $\Delta$M is the mass lost during the red giant  phase. We assume that $\Delta$M
has a gaussian dispersion $\sigma$ around an average value $\Delta$M$_{0}$ and
that both $\Delta$M$_{0}$ and  $\sigma$ are parameters to be determined and do not
depend on Y. M$_{\rm RG}$ decreases with increasing helium content, so that the stars with higher
helium will have smaller H-envelope mass. Fixed $\Delta$M$_{0}$ and $\sigma$ we extract random both the mass loss and the HB age in the interval from $10^6$ to $10^8$ yr. We thus locate the luminosity and Teff along
the evolution of the HB mass obtained. 
The L and Teff values, are then  transformed into the different observational
magnitudes using the synthetic  color transformations to ACS bands by \citet{dotter2007}.\\
Population synthesis for the main sequence and the SGB follows the lines described in \cite{ventura2009}.
\\
\begin{figure*}
\centering
\includegraphics[width=15cm]{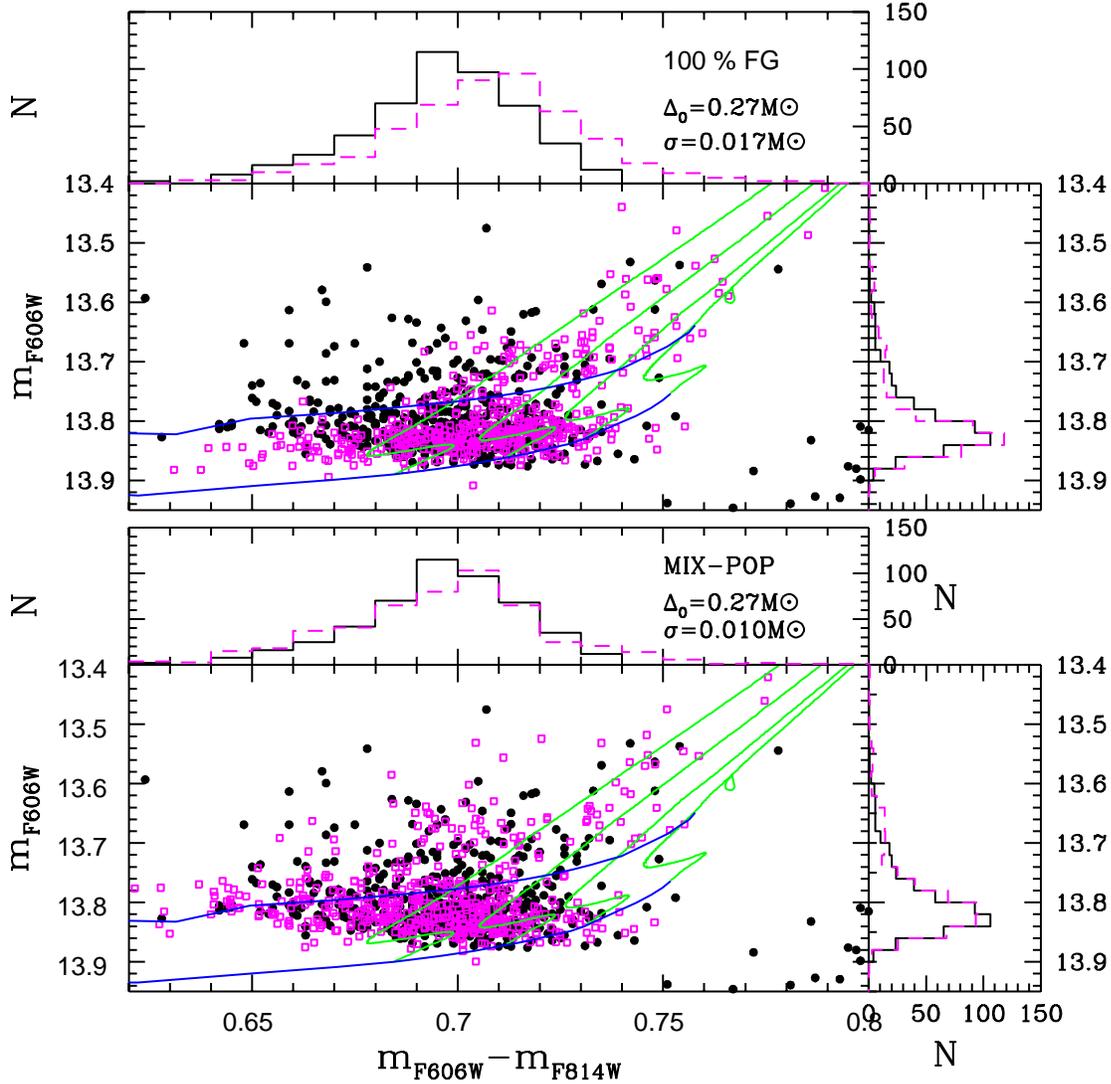}
\vspace{-80pt}
\caption{Comparison  between observed (filled circle) and  two different synthetic HB simulations (open squares-magenta in the electronic version of the paper) Upper panels: synthetic population is built under the hypothesis that a unique generation of stars with Y=0.25 is present. Even using a large value of $\sigma$ it is not possible to  reproduce at the same time the spread in color and magnitude. We also show the HB evolutionary tracks of stars
with masses M=0.63, 0.66, 0.70 and 0.80 solar masses and the ZAHBs for normal populations with respectively Y=0.25 (low luminosity) and Y=0.28 (high luminosity) shifted as described in the caption of Fig.1.
Lower panel: The synthetic population is  built under the hypothesis that 
70 \% of stars has Y randomly chosen between Y=0.25 and Y=YUP=0.27 (see text). 
The  dashed histograms refer to synthetic populations, while the solid ones are those relative to observations.}
\label{fig2}
\end{figure*}
\section{The results}
The idea we want to test in this work is that the two separated SGBs found by \citet{anderson2009} are due to the presence of two stellar populations with different  initial  mixtures of elements heavier than helium, mainly C+N+O. In particular,  we explore the possibility that CN strong stars all belong to the SG, and were formed from gas showing the signature of CNO processing, and a helium enrichment, but that only a small percentage of SG stars are increased in the overall CNO content.\\
Following the suggestion of \citet{dantonacaloi2008} we decide first to investigate this hypothesis using  the morphology of the HB of 47 Tuc, where a population with a larger  helium abundance should emerge, and then test the results using the data of the SGB by \citet{anderson2009}.   
First we review previous literature concerning the HB of 47 Tuc and the features of metal rich
HBs.
\subsection{Previous analysis of the HB data for 47 Tuc and of the HB morphology of metal rich clusters}
Already \citet{dorman1989} attempted to derive information on the evolution in the HB of 47Tuc, 
by comparing the (B,V) data by \citet{hesser1987} ($\sim$50 HB stars) with  
evolutionary models  computed for different chemistry. This work shows that the appropriate
helium content for the gas out of which these stars formed had $\sim$24$\%$ helium by mass, 
a value consistent with the (previous) estimates of primordial helium abundance.
No population synthesis was attempted, and in any case the sample was too small to draw any
inference on possible helium spread among the stars in this HB but they argued for the first time  that the bluest HB stars
 should be somewhat brighter than the ZAHB, due to evolution. Population synthesis for metal
rich compositions was extensively used by \cite{catfp1996} who show that some clump simulations have a 
wedge--shaped structure, due to the population of the long loops of the evolutionary tracks they
adopt, but no quantitative comparison with 47Tuc data is available. 
Other works have studied the tilted morphology of several metal rich clusters
attributing it as a natural outcome of standard evolutionary theories 
\citep[for example see the case of NGC 6362 in][]{brocato1999}, or the consequence of the 
presence of a differential redding \citep{raimondo2002}. Anyway, 47Tuc does not fall 
among the clusters having a ``tilted" clump, as shown, for example, in the Fig.9 
of the last mentioned paper. 
In fact, \citet{carney1993} suggested that the 47Tuc HB is ``wedge-shaped", based on the data by Hesser et
al. 1977. 

\subsection{A new analysis of the Horizontal Branch}
In the top panel of  Fig. \ref{fig1} we report a zoom of CMD of 47 Tuc at HB level, including $\sim$500 stars, and based on images of 5sec  in F606W and F814W from ACS (GO-10775, PI Sarajedini). 
The photometric  reduction of  the ACS/WFC data  was  carried out
using the software  presented and described in detail  in \citet{anderson2008}. It consists in  a package
which allowed us to analyze all the exposures simultaneously to generate a single star list.  Stars
are measured independently in each image by using a spatially varying
9$\times$10 array of empirical ``library PSFs'' from \citet{anderson2006}, plus a spatially constant perturbation for each exposure, to
account for variations in the telescope focus. This routine was  designed to work well in  both crowded and uncrowded
fields, and it is able to detect almost every star that can be detected
by eye.  Calibration of  ACS photometry into the  Vega-mag system
was performed following recipes in \citet{bedin2005} and using
the zero points given in \citet{sirianni2005}.
Unfortunately, the hybrid PSF model above is not able to account for all
of the effects of telescope breathing, which can introduce a small
spatial dependence of the shape of the PSF, which is not compensated for
in our PSF model and can cause small systematic photometric errors that
depend on position on
the detector.  The typical variation is small (about 1\% in the fraction
of light in the core).  To account for these variations, we used the
method adopted in \citet{milone2008}.\\
Fig. \ref{fig1} shows the HR diagram. As the photometric errors are very small ($\sim$ 0.01 mag on each filter), we can see that the
``wedge-shaped" morphology is in fact reduced to the presence of a small dent in luminosity, at color
m$_{\rm F606W}$-m$_{\rm F814W}$ $\sim$0.675 mag, between the blue (and brighter) side and 
the red (and dimmer) side. Figure 8 in \cite{briley1997} shows that the brighter HB stars (including the bluer ones) are preferentially CN strong, while the dimmer ones are all CN weak. This is the strongest hint that we are looking at a population effect and not at an evolutionary effect as first  suggested by \citet{dorman1989}\\
 The theoretical zero age HB (ZAHB) for two different values of Y (0.25 and 0.28) are shown, reported on the observational plane using  a reddening E(606-814)=0.038 mag, corresponding to  the E(B-V)=0.04 mag given in literature according to the relationships by \citet{bedin2005}.
The apparent distance modulus (m--M)$_{\rm F606W}$=13.09 mag was chosen  in order to fit a ZAHB of pristine Y(=0.25)  to the lower envelope of the observed HB.\\
We see from Fig.1 that the ZAHB with Y=0.25 provides a good fit
of the reddest stars observed, whereas the bluer part is not fit  and appears to require a larger helium.
This supports our idea that a SG of stars is present in 47 Tuc, with an higher abundance of helium.
In Fig. 1 we also show the ZAHBs with CNO$\uparrow$(dashed line)
which are  little fainter than
the ZAHBs corresponding to the normal CNO abundance, in agreement with
the fact that the total metallicity of the new mixture is a little higher.
\begin{figure*}
\centering
\includegraphics[width=12cm]{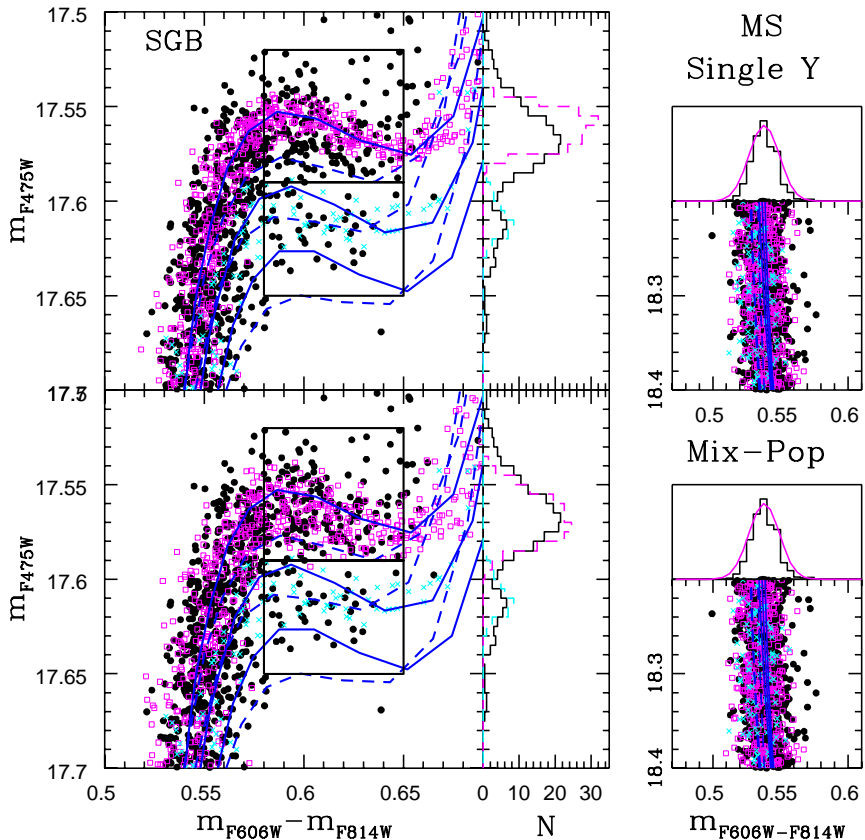} 
\vspace{-80pt}
\caption{The same as Fig. \ref{fig2} but in this case the SGB region is shown. Crosses (cyan  color in the electronic version) are used to mark those 10 \% of simulated stars obtained from CNO$\uparrow$ models  using the same N(Y) and $\Delta$M used for  the rest of population (see text). 13 Gyr isochrones calculated  from normal and enhanced C+N+O models with primordial helium abundance (solid lines) and  Y=0.28 (dashed lines) are shown. Down from bright luminosity we have respectively CNO normal, CNO$\uparrow$, CNO$\uparrow$$\uparrow$ isochrones. Histograms show the comparison between observed and simulated distributions of magnitude of stars confined in two box drawn in the CMD in order to isolate the two components of SGB. It is evident  from the lower panels that the case of a mixed population with a 70 \% of total stars belonging to a SG,  a part of which ($\sim$10\%) have also a little increase of C+N+O elements, fit better the observations. On the right panels the comparison is shown at the level of MS for the same set of data, in order to explain the method used to choice the photometrical errors included in our simultations ($\sim$0.008mag, see text).}
\label{fig3}
\end{figure*}

To compute a synthetic population of the HB of 47 Tuc we adjusted the mean mass loss on the  
RGB and its dispersion in order to reproduce the feature  we attribute to the FG, 
in particular the dent described before. We assume that 30\% of the stars (the FG) 
formed with the primordial helium abundance Y=0.25. We find that the best choice is 
$\Delta$M$_{0}$=0.27M$_{\odot}$ and $\sigma$=0.010M$_{\odot}$ for an age of 12 Gyr.
The same distribution of mass loss is used for the remaining 70\% of stars, a population 
with Y randomly distributed between Y=0.25 and Y=YUP, and YUP is  chosen to reproduce 
both the color and magnitude distributions of the observed HB.
The choice of 30\% of stars for the FG is consistent with 
both the CN strength sample \citep[e.g][]{briley1997} and the [Na/O] anticorrelation 
sample \citep{carretta2009}.\\
Fig.2 shows that the choice YUP=0.27, coupled with the same distribution of mass
loss assumed for the FG stars, allows to nicely reproduce both the color and
magnitude distributions observed on the HB.\\
On the other hand, any attempt to model the observed HB with a single
population proved much less satisfactory, independently of the choices concerning mass
and the distance modulus. We show as an example (upper panel of Fig.2)
the synthetic population obtained when a larger dispersion in the mass loss 
($\sigma$=0.017 M$_{\odot}$): an acceptable fit of the color dispersion is accompanied by a
poorer match of the magnitude spread.
Tracks having loops more extended towards the blue, for a given chemical composition, would reduce the value of 
mass dispersion, but leave the problem unchanged. Of course such a detail of
behavior would remain hidden if the observations were of lower quality. In addition, the spectroscopic
confirmation that the bluer and brighter stars are CN strong (and thus belonging to the SG, and
possibly helium rich) is fundamental to give weight to this interpretation.  
\subsection{Sub  Giant Branch}
In Fig.3 we show the SGB and a small portion of the  MS  of the inner region of 47 Tuc observed with WFC at HST's Advanced Camera for Surveys. Two archive data set used here are :1) GO-9028 for F475W photometrical band (20X60s)  and 2) GO-10775 for F606W (4x50 s) and F814W (5x50 s). The reduction strategy is  described in \citet{anderson2009}. In the figure also 13 Gyr isochrones calculated  from normal and enhanced C+N+O models with primordial helium abundance (solid lines) and  Y=0.28 (dashed lines) are shown. Down from bright luminosity we have respectively CNO normal, CNO$\uparrow$ and models without the dilution described in previous section (CNO$\uparrow$$\uparrow$). It is important to note that for a fixed age and C+N+O content, the luminosity  of SGB decreases by $\sim$ 0.02 mag by increasing  the initial helium abundance by $\Delta$Y=0.03\footnote{This result complements previous analysis \citep[see for example][]{ventura2009, catelan2010}, stating that the position of the SGB is not sensitive to Y. The previous results where based on metal poor models. The dependence of the SGB location on Y becomes important only in metal rich models.}. We attribute the spread of the SGB  of 47 Tuc found by \citet{anderson2009} to this effect.\\
We adopt the same N(Y)  distribution inferred to reproduce the  HB,  to simulate the TO and SGB and in Fig. \ref{fig3} (lower panels )we show the results.\\
In our simultations we use as  photometric errors those which reproduce the widht of the main sequence obtained from the same set of data as shown in the right panels in Fig.3  since, as shown in \citet{anderson2009} from  the analysis of a less crowed region 6' from the center of 47 Tuc, the intrinsic breadth of  the MS is much more narrow ($\Delta$(m$_{F606W}$-m$_{F8814W}$)$\sim$0.010mag). \\
In this way we can  reproduce the spread of the bright SGB and show that a  simple population cannot do the same good job. This is shown in the upper pannels of Fig. \ref{fig3} where a  a single abundance of Y is considered with the right choice of the photometrical errors to reproduce the width of MS.\\
Finally to complete the analysis and to explain  also the faint component of SGB we make the assumption that  10\% of the total population  belongs to a  SG  and has an higher CNO abundance (CNO$\uparrow$, crosses in Fig.\ref{fig3} ). This component is not recognisable in the HB, since ZAHBs calculated for a normal mixture and CNO$\uparrow$ are substantially the same (Fig.\ref{fig1}).\\


\section{Discussions and conclusions}
Our simulations show that the morphology the HB of 47 Tuc needs a non negligible spread in the helium abundance ($\Delta$Y=0.02) to be explained, as originally suggested by \citet{briley1997}.
This spread is  consistent with the value  suggested by \citet{anderson2009} from their  analysis of the width of the MS ($\Delta$Y=0.027) especially if one take into account that the last one is obtained in the hypothesis that all the MS spread is due to helium, while there can be other color effects as suggested by the authors in their paper.
The most interesting result is that  the same variation  in helium can also explain  the spread of the bright SGB while the faint SGB  is made of stars with  higher C+N+O.\\ We interpret these results as the confirmation that SG in 47 Tuc consist   of about 70 \% of stars, as suggested by spectroscopic studies.
Summarizing 47 Tuc is made up of three different sub-popolations:
\begin{enumerate}
\item {\bf FG}; it consists in 30\% of the cluster stars that we have recognized as first generation stars. In the CMD they are located on the red part of the HB and form the extreme upper part of the bright SGB.
\item {\bf SGI}; formed by the  60\% of stars belonging to  a  second generation  characterized by the same C+N+O of FG but having a dispersion of Y between the primordial value (Y=0.25) e YUP=0.27. This population evolves in the bright part of the HB, including the bluer stars of the ''step'' and is responsible  for  the spread of the bright SGB. If the polluting matter is identified with the CNO processed gas ejected by AGB stars \citep{ventura2001}  this population formed from material ejected from AGB progenitors so massive ($>$5 M$_{\odot}$)  that the chemistry of the ejecta is scarcely affected by third dredge up.
\item {\bf SGII}; it is made up  by  10\% of stars wich are C+N+O enhanced, and emerge as faint SGB in the CMD but are  not recognizable in the HB. We interpret this population as made up of stars born from material C+N+O enhanced by a factor 1.4  and diluted by 50 \% with pristine material (mixture CNO$\uparrow$). The presence of the faint SGB gives  a lower limit to the AGB mass which have contributed to the formation of SG giving  an indication on how long the phase of formation of SG last. Following our interpretation this time is $\sim$ $10^8$ yr after the formation of FG. Obviously this result depends on the dilution model used and on the models from which was calculated chemical  yields.
However the spread we can see among the luminosities of the few stars  belonging to the dimmer SGB (see Fig. 3) may also indicate an additional helium or C+N+O spread among the stars of this small population.
\end{enumerate}
Concluding, we remember that, as shown in the hydrodynamic plus N--body simulations of the SG  formation in globular clusters by \citet{dercole2008} a larger concentration of these SG stars can  be expected, in some cases at the center of the cluster. This  result is compatible with our interpratation since  \citet{briley1997} has shown that  in the inner part of the cluster there is a higher fraction of stars with strong CN absorption (which we interpret as SG stars)
than in the outer parts.\\
\section*{Acknowledgments}
We thank A. Dotter for providing the  color-T$_{\rm eff}$  transformations for ACS bands, M.Marconi for useful discussions  and  L.Bedin and J.Anderson for  providing us  the data used in this article. F.D. thanks Gary Da Costa for providing information of the work by G. Paltoglou.  Financial support for this study was provided by the PRIN MIUR 2007 ``Multiple stellar populations in globular clusters: census, characterization and origin''.
{}
\end{document}